\documentclass[prl,twocolumn]{revtex4}
\usepackage{epsf}
\usepackage{psfig}
\usepackage{graphicx}
\begin{document}


\title{Large negative magnetoresistance in a ferromagnetic shape memory alloy : Ni$_{2+x}$Mn$_{1-x}$Ga} 

\author{C. Biswas, R. Rawat, and S. R. Barman}
\affiliation{UGC-DAE Consortium for Scientific Research,
Khandwa Road, Indore, 452017, Madhya Pradesh, India.}

\begin{abstract}  
5\% negative magnetoresistance (MR) at room temperature has been observed in bulk Ni$_{2+x}$Mn$_{1-x}$Ga. This indicates the possibility of using Ni$_{2+x}$Mn$_{1-x}$Ga as magnetic sensors. We have measured MR in the ferromagnetic state for different compositions ($x$=0~-~0.2) in the austenitic, pre-martensitic and martensitic phases. MR is found to increase with $x$. While MR for $x$=0 varies almost linearly in the austenitic and pre-martensitic phases, in the martensitic phase it shows a cusp-like shape. This has been explained by the changes in twin and domain structures in the martensitic phase. In the austenitic phase, which does not have twin structure, MR agrees with theory based on s-d scattering model. 
\end{abstract}

\maketitle

Ni$_{2+x}$Mn$_{1-x}$Ga is a technologically  important material for its potential application as a magnetically driven shape memory alloy (SMA) that is more efficient than temperature or stress driven SMA devices.[1-3] So, in recent years, there is a flurry of activity in this field.[1-24]  
Magnetic field induced strain  up to 9.5\% has been reported in Ni$_2$MnGa.\cite{Sozinov}   
It has been reported that the Curie temperature ($T_C$) and martensitic transition temperature ($T_M$) of  Ni$_{2+x}$Mn$_{1-x}$Ga can be varied  systematically with composition.\cite{Vasilev99,Zuo99} For example, varying $x$ from 0 to 0.2 causes  $T_C$ to decrease from 376 to 325~K and $T_M$ to increase from 210 to 325~K.  Giant-magnetocaloric effect has been observed in samples where $T_C$ and $T_M$ are equal.\cite{Marcos03,Zhou04} 

In this paper, we report 5\% negative magnetoresistance (MR) at 8~T in bulk Ni$_{2+x}$Mn$_{1-x}$Ga polycrystals at room temperature.  To the best of our knowledge, there are no MR studies in literature till date on well characterized bulk Ni$_{2+x}$Mn$_{1-x}$Ga samples as functions of composition and temperature. Lund et. al.\cite{Lund02} reported MR of Ni$_2$MnGe and Ni$_2$MnGa films on GaAs(001) to be $\approx$~1\% at 9~T at 280~K. Non-stoichiometric Cu-Al-Mn shape memory Heusler alloy shows a large negative MR of 7\%  at 5~T at 10~K, but at 250~K it is only 0.5\%.\cite{Marcos02} In half-Heusler alloys like NiMnSb and PtMnSb, the reported MR values at 295~K and 8~T are $\approx$1\% and $\approx$2.5\%, respectively and are attributed to the inelastic s-d scattering.\cite{Moodera} 

The polycrystalline ingots of Ni$_{2+x}$Mn$_{1-x}$Ga for $x$ = 0, 0.1 and 0.2 were prepared by standard method\cite{Vasilev99} and have been characterized by x-ray diffraction (XRD), energy dispersive x-ray analysis 
(EDAX), differential scanning calorimetry (DSC), resistivity, ac magnetic susceptibility, magnetization and photoelectron spectroscopy. The structure and lattice constants 
from Rietveld refinement of the XRD pattern agree with Refs.\cite{Webster84,Wedel99}. DSC, resistivity, ac susceptibility and magnetization measurements are in agreement with literature.\cite{Hu,Vasilev99,Zuo99} From these studies we find the $T_M$ (matensitic start temperature) and $T_C$ of Ni$_2$MnGa to be 207 and 365~K, respectively.  EDAX shows that the samples are homogeneous and the intended
and actual compositions agree well, {\it e.g.} Ni$_{2.02}$Mn$_{0.97}$Ga$_{1.02}$,
and Ni$_{2.21}$Mn$_{0.78}$Ga$_{1.01}$ for $x$=0 and 0.2, respectively. The photoelectron spectra are in agreement with theory.\cite{Chakrabarti04}   Magnetization and ac-susceptibility measurements clearly show that the samples are ferromagnetic at and below room temperature.  The isothermal MR measurements have
been done in the ferromagnetic state at three different temperatures corresponding to the tetragonal martensitic, pre-martensitic and cubic austenitic phases. For measurements at lower temperatures, the 
samples have been cooled from 300~K in zero field. The MR is calculated using the standard expression :  $\Delta$$\rho$/$\rho$$_0$ = ($\rho$$_H$~-~$\rho$$_0$)/$\rho$$_0$, where $\rho$$_0$ is the resistivity at zero field.

\begin{figure}[htb]
\epsfxsize=80mm
\epsffile{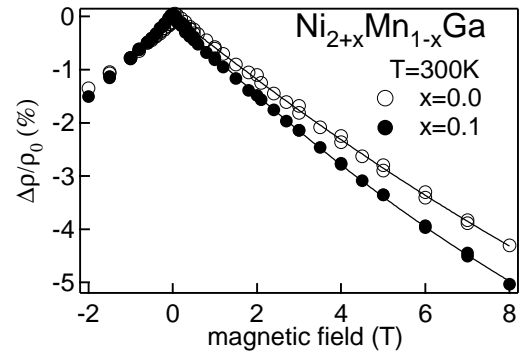}
\caption{Isothermal magnetoresistance (MR) curves as a function of magnetic field 
for Ni$_{2+x}$Mn$_{1-x}$Ga with $x$=0 and 0.1 in ferromagnetic state at 300~K. The solid lines are fit to the data.}
\label{fig1}
\end{figure}

In Fig.~1, we show MR of Ni$_{2+x}$Mn$_{1-x}$Ga for $x$=0 and 0.1 at 300~K, where both the samples are ferromagnetic.  The variation of MR is almost linear, which we have fitted with a second order polynomial (solid line). The deviation from linearity is quantified by the ratio of the magnitudes of the second order and the linear term (0.02 in both cases).  Ni$_{2+x}$Mn$_{1-x}$Ga is an ideal local moment ferromagnet where the magnetic moment is mainly localized on Mn ions ($\approx$3.84$\mu$$_B$) and Ni atoms contribute small magnetic moment ($\approx$0.33$\mu$$_B$).\cite{Webster84,Chakrabarti04} MR for stable ferromagnets with localized moments and high carrier concentration has been recently calculated by Kataoka on the basis of $s-d$ model, where $s$ conduction electrons are scattered by localized $d$ spins.\cite{Kataoka01} The author has considered the difference in the relaxation time of the majority- and minority-spin carriers due to the splitting of spin bands. The calculated MR variation below $T_C$ (in our case $T$=300~K, and $T_C$ for $x$=0 is 365~K, giving $T$/$T_C$=0.8) is in very good agreement with the present experimental results (see for example Fig.~8 of Ref.\cite{Kataoka01}). From Fig.~1, we also observe that for $x$=0 MR at 8~T is 4.3\%, while for $x$=0.1 it increases to 5\%. We note that $T_C$ decreases from 365~K to 350~K between $x$=0 and 0.1. Thus, $T_C$ for $x$=0.1 is closer to the measurement temperature (300~K). It is well known that as T approaches $T_C$, MR increases due to magnetic spin disorder scattering.\cite{Lund02,Kataoka01} Thus, decrease in $T_C$ with increasing $x$ results in enhanced MR at room temperature. 
Lund {\it et al.} have not observed such large MR in Ni$_2$MnGa thin films probably because of compositional differences or granularity as evidenced by absence of martensitic transition and decreased $T_C$ compared to the bulk, or other reasons like interface reaction, substrate or intrinsic effects.

\begin{figure}[htb]
\epsfxsize=80mm
\epsffile{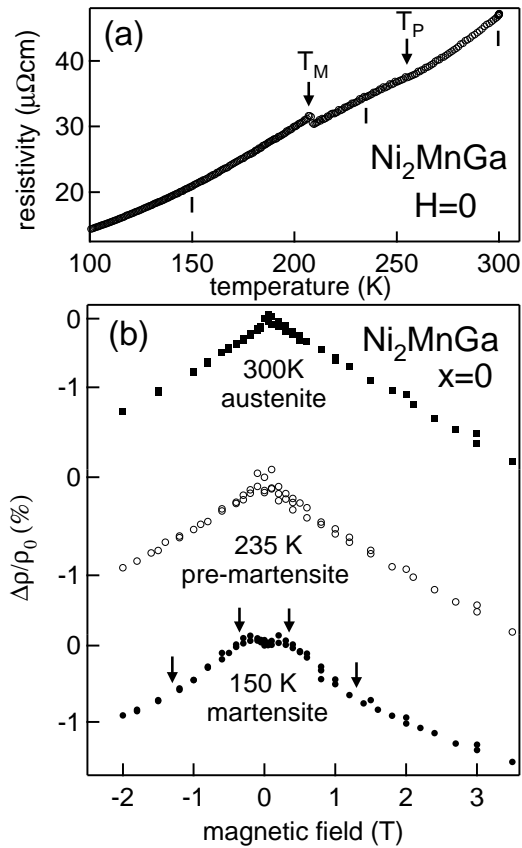}
\caption{(a) Resistivity of Ni$_2$MnGa ($x$=0) at zero field. The temperatures for MR measurement are shown by ticks. (b) MR of Ni$_2$MnGa for the different phases in the ferromagnetic state (T$_C$=365~K). The arrows indicate the points of inflection.} 
\label{fig2}
\end{figure}

In Fig.~2a, we show the zero field resistivity of Ni$_2$MnGa as a function of temperature. Resistivity exhibits a metallic behavior in agreement with literature\cite{Vasilev99,Zuo99} with T$_M$~=~207~K and a pre-martensitic  transition at $T_P$~=~255$\pm$5~K, as expected for the $x$=0 stoichiometric composition.\cite{Zuo99} Fig.~2b shows MR of Ni$_2$MnGa in the austenitic (300~K), pre-martensitic (235~K) and the martensitic (150~K) phases, which are all in ferromagnetic state. The behavior of MR between austenitic and martensitic phases in the low field region is different. In the martensitic phase, MR exhibits a cusp-like shape below 1.3~T with two points of inflection (shown by arrows). Between 0 and 0.3~T MR hardly varies, while between 0.3 and 1.3~T there is a substantial increase and above 1.3~T, MR increases gradually.  

In order to understand the above observations, we discuss the differences in the structural and magnetic properties of the martensitic and austenitic phases. The austenitic phase has an L2$_1$ cubic structure which upon martensitic transition changes to a tetragonal structure. In the martensitic phase, twinning takes place with different twin variants to reduce the strain.  Although the magnetic moments are not considerably different, the magnetocrystalline anisotropy constant ($K_1$) is large for the martensitic phase (3$\times$10$^6$ erg/cm$^3$), whereas it is very small in the austenitic phase.\cite{Albertini01} So, the magnetization saturates rapidly in the austenitic phase in contrast to the martensitic phase where the change is gradual.\cite{Albertini01} 
Due to twinning and large $K_1$, the effect of magnetic field in the martensitic phase is more complicated resulting in twin-boundary motion and variant nucleation. Each variant has large $K_1$ with the easy axis along [001] direction giving rise to the rearrangement of the twin related variants under the driving force originating from the difference in the Zeeman energy of the different variants. When $K_1$ is larger than  Zeeman energy, the effect of magnetic field is to move the twin boundaries rather than magnetization rotation within the unfavorably oriented twins.\cite{Handley} Hence, the saturation magnetization is achieved mainly by twin-boundary motion. From magnetic force microscopy measurements on single crystal Ni$_{2.05}$Mn$_{0.96}$Ga$_{0.99}$ with field along [010] direction, Pan {\it et al.}\cite{Pan00} found that between 0-0.2~T both twin boundary and domain wall motion take place giving rise to a spike domain structure with reverse domains. However, for a polycrystal this effect might be negligible due to random orientation of the grains causing hardly any change in MR in the low field region (0-0.3~T). Above 0.2~T, the reversed magnetic domains start disappearing, domain wall motion and magnetization rotation take place  gradually changing  the whole twin to a single domain and finally around 1~T the sample is very close to a single variant state.\cite{Pan00} 
Correspondingly, in the region between 0.3 to 1.3~T we find MR to increase substantially. So, this can be related to the saturation of sample magnetization resulting in suppression  of  magnetic spin-disorder scattering.\cite{Lund02,Kataoka01} Thus, MR in the martensitic phase shows different behavior compared to the austenitic phase due to twinning and high $K_1$. On the other hand, the pre-martensitic phase MR (Fig.~2b) is similar to the austenitic phase because it is essentially the austenitic phase with a micro-modulated structure and no twinning.\cite{Manosa01} 

\begin{figure}[htb]
\epsfxsize=80mm
\epsffile{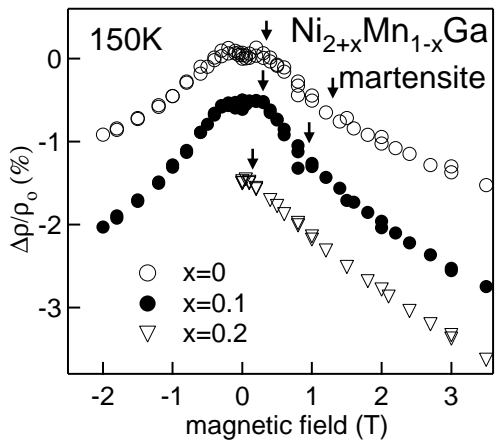}
\caption{MR of Ni$_{2+x}$Mn$_{1-x}$Ga ($x$=0, 0.1 and 0.2) in the martensitic phase. The arrows indicate the points of inflection. The curve for x=0.1 and x=0.2 are staggered by -0.5\% and -1.5\% respectively for the clarity of presentation.} 
\label{fig3}
\end{figure}

In Fig.~3, we show MR of ferromagnetic Ni$_{2+x}$Mn$_{1-x}$Ga for $x$=0, 0.1 and 0.2 in the martensitic phase at 150~K. We find that MR changes with composition. For $x$=0.1, the inflection points (shown by arrows) are observed at lower H compared to $x$=0. In contrast, $x$=0.2 is almost linear with a possible inflection point at 0.15~T. It has been experimentally found that K$_1$ decreases by about 55\% between $x$=0 and 0.15.\cite{Albertini02} The decrease in K$_1$ implies that domain wall motion and magnetization rotation would dominate over twin boundary motion. Hence, the normal MR behavior for a ferromagnet\cite{Kataoka01} would gradually emerge with higher Ni doping. 

In conclusion, we report large negative magnetoresistance (5\% at 8~T) for bulk polycrystalline Ni$_{2+x}$Mn$_{1-x}$Ga, which is highest reported so far for a shape memory alloy at room temperature. This opens up  new prospects of  technological application for this shape memory alloy. 

We thank Prof. M. Kataoka for useful discussions. Mr. M. Manekar, Dr. N. P. Lalla, Dr. A. M. Awasthi, Dr. A. Banerjee, Dr. S. B. Roy, Dr. D. M. Phase, and Dr. C. Mukhopadhyay are thanked for sample characterization. Prof. V. N. Bhoraskar and Prof. A. Gupta are thanked for support.\\


\end{document}